\documentclass[aps,prd,twocolumn,superscriptaddress,nofootinbib]{revtex4-1}

\usepackage{latexsym}
\usepackage{amsmath}
\usepackage{amssymb}
\usepackage{amsfonts}
\usepackage{bm}
\RequirePackage{lineno}

\usepackage{color}
\definecolor{purple}{rgb}{0.5,0,0.5}
\definecolor{blue}{rgb}{0.0,0,0.9}
\definecolor{prdblue}{rgb}{0.133,0.118,0.498}
\usepackage[colorlinks=true, pdfstartview=FitV, linkcolor=prdblue, citecolor= prdblue, urlcolor=prdblue]{hyperref}

\usepackage{supertabular} 
\usepackage{placeins}
\usepackage{epsfig}
\usepackage{graphicx}
\usepackage{hyperref}

\hypersetup{
  breaklinks=true,
  colorlinks = true,
  linkcolor = blue,
  anchorcolor = blue,
  citecolor = blue,
  filecolor = blue,
  pagecolor = blue,
  urlcolor = blue
}
\hypersetup{
  bookmarks=true,
  bookmarksnumbered=true,
  bookmarkstype=toc,
  linktocpage=true
}

\begin{document}

\modulolinenumbers[2]

\setlength{\oddsidemargin}{-0.5cm} \addtolength{\topmargin}{15mm}

\title{\boldmath First observation of the semileptonic decay $\Lambda_c^+\rightarrow pK^- e^+\nu_e$}
\author{
\small
M.~Ablikim$^{1}$, M.~N.~Achasov$^{11,b}$, P.~Adlarson$^{70}$, M.~Albrecht$^{4}$, R.~Aliberti$^{31}$, A.~Amoroso$^{69A,69C}$, M.~R.~An$^{35}$, Q.~An$^{66,53}$, X.~H.~Bai$^{61}$, Y.~Bai$^{52}$
% ...
\\
\vspace{0.2cm}
(BESIII Collaboration)\\
\vspace{0.2cm} {\it
$^{1}$ Institute of High Energy Physics, Beijing 100049, People's Republic of China\\
% ...
$^{4}$ Bochum Ruhr-University, D-44780 Bochum, Germany\\
% ...
$^{11}$ G.I. Budker Institute of Nuclear Physics SB RAS (BINP), Novosibirsk 630090, Russia\\
% ...
$^{31}$ Johannes Gutenberg University of Mainz, Johann-Joachim-Becher-Weg 45, D-55099 Mainz, Germany\\
% ...
$^{35}$ Liaoning Normal University, Dalian 116029, People's Republic of China\\
% ...
$^{52}$ Southeast University, Nanjing 211100, People's Republic of China\\
$^{53}$ State Key Laboratory of Particle Detection and Electronics, Beijing 100049, Hefei 230026, People's Republic of China\\
% ...
$^{61}$ University of Jinan, Jinan 250022, People's Republic of China\\
% ...
$^{66}$ University of Science and Technology of China, Hefei 230026, People's Republic of China\\
% ...
$^{69}$ University of Turin and INFN, (A)University of Turin, I-10125, Turin, Italy; (B)University of Eastern Piedmont, I-15121, Alessandria, Italy; (C)INFN, I-10125, Turin, Italy\\
$^{70}$ Uppsala University, Box 516, SE-75120 Uppsala, Sweden\\
% ...
$^{b}$ Also at the Novosibirsk State University, Novosibirsk, 630090, Russia\\
% ...
\vspace{0.4cm}
}
}
\begin{abstract}
Using $4.5~\mathrm{fb}^{-1}$ of $e^+e^-$ annihilation data samples collected at the center-of-mass energies ranging from 4.600~GeV to
4.699~GeV with the BESIII detector at the BEPCII collider, a first study of the semileptonic decays $\Lambda_c^+\rightarrow pK^-e^+\nu_e$, $\Lambda_c^+\rightarrow \Lambda(1520) e^+\nu_e$ and $\Lambda_c^+\rightarrow \Lambda(1405) e^+\nu_e$ is performed. The $\Lambda_c^+\rightarrow pK^-e^+\nu_e$ decay is observed with a significance of $8.2\sigma$ and the branching fraction is measured to be $\mathcal{B}(\Lambda_c^+\rightarrow pK^- e^+\nu_e)=(0.88\pm0.17_{\rm stat.}\pm0.07_{\rm syst.})\times 10^{-3}$. 
We also report evidence of $\Lambda_c^+\rightarrow \Lambda(1520)e^+\nu_e$ and $\Lambda_c^+\rightarrow \Lambda(1405)e^+\nu_e$ with significances of $3.3\sigma$ and $3.2\sigma$, respectively, and measure $\mathcal B(\Lambda^+_c\rightarrow \Lambda(1520)e^+\nu_e)=(1.02\pm0.52_{\rm stat.}\pm0.11_{\rm syst.})\times10^{-3}$ and $\mathcal B(\Lambda^+_c\rightarrow \Lambda(1405)[\rightarrow pK^-]e^+\nu_e)=(0.42\pm0.19_{\rm stat.}\pm0.04_{\rm syst.})\times10^{-3}$. 
Combining these with the inclusive semileptonic $\Lambda_c^+$ branching fraction measured by BESIII, the relative fraction is determined to be 
$[\mathcal{B}(\Lambda_c^+\rightarrow pK^-e^+\nu_e)/\mathcal{B}(\Lambda_c^+\rightarrow X e^+\nu_e)]=(2.1\pm0.4_{\rm stat.}\pm0.2_{\rm syst.})\%$,
which provides a clear confirmation that semileptonic $\Lambda_c^+$ decays are not saturated by the $\Lambda \ell^+\nu_{\ell}$ final state.
\end{abstract}

\maketitle

The study of $\Lambda^+_c$ semileptonic\,(SL) decays provides valuable
informations about weak and strong interactions in baryons containing a 
heavy quark.  (Throughout this paper, charge-conjugate modes are
implied unless explicitly noted.) 
Their decay rates depend on the weak quark mixing Cabibbo-Kobayashi-Maskawa (CKM) matrix~\cite{CKM} element $|V_{cs}|$ and strong interaction effects parameterized by form factors describing the hadronic transition between the initial and the final baryons. 
In comparison to experimental studies of SL decays in the charmed meson sector~\cite{pdg2020}, rather few measurements exist of $\Lambda_c^+$ SL decays. 
No other exclusive SL decay except $\Lambda_c^+\rightarrow \Lambda \ell^+\nu_{\ell}$ ($\ell=e, \mu$) has been reported to date~\cite{PRL115_221805,PLB767_42}. In addition, a comparison of the branching fractions (BFs) for the exclusive decay $\Lambda^+_c\rightarrow \Lambda \ell^+\nu_{\ell}$ 
and inclusive decay $\Lambda_c^+\rightarrow X\ell^+\nu_{\ell}$ shows that their ratio is close to one~\cite{PRL121_251801}, which exhibits a different pattern compared with charm mesons~\cite{pdg2020,PRD104_013005,2103.07064}. 
For example, the BF of $D^{+(0)}\rightarrow \bar{K}^0(K^-)e^+\nu_e$ is much smaller than $\mathcal{B}(D^{+(0)}\rightarrow e^+X)$~\cite{pdg2020}.
Searching for unknown exclusive SL $\Lambda_c^+$ decay can provide important information to validate and understand this pattern.
The decay $\Lambda_c^+\rightarrow pK^-e^+\nu_e$ is one of the best suited channels to search for~\cite{PRC72_035201,PRD93_014021,PRD95_053005,PRD97_116015,PRD105_L051505,PRD105_054511}, as its final state is simple with a high detection efficiency, in comparison to decays such as $\Lambda_c^+\rightarrow \Sigma\pi e^+\nu_e$ etc. 

Since the $ud$ diquark is a spectator, the SL $\Lambda_c^+\rightarrow pK^-e^+\nu_e$ decay provides a perfect filter of isospin $I=0$ meson-baryon states with almost no contamination from the $I=1$ amplitude~\cite{PRD93_014021,PRC92_055204}. This provides an ideal platform to study the internal structure of exotic $\Lambda^*$ states. 
Among these states, particular interest is concentrated on the $\Lambda(1405)$, in which the high-mass pole strongly couples to $\bar{K}N$ final-state~\cite{pdg2020}. 
The $\Lambda(1405)$ is considered as the most striking state in understanding the spectrum of baryons with strangeness and has been continuously studied for more than 60 years since its theoretical prediction~\cite{PRL2_425,AnnPhys}.
However, the nature of $\Lambda(1405)$ is still mysterious~\cite{Prog120,EPJST}. It is suggested to be a dynamically generated meson-baryon molecular state~\cite{AnnPhys,PR153_1617,PRD93_014021} or a three-quark $uds$ bound state~\cite{PRC72_035201,PRD95_053005}. 
The decay of $\Lambda_c^+\rightarrow \Lambda(1405)e^+\nu_e$ is expected to be a promising process to distinguish its structure because the predicted BF of the decay in the two hypotheses differ by a factor of roughly 100 times~\cite{PRC72_035201,PRD93_014021,PRD95_053005}, as shown in Table~\ref{tab:theorybf} discussed later.

Furthermore, in heavy-baryon SL decays, most previous LQCD calculations are concentrated on the transition of $J^P=1/2^+\rightarrow J^P=1/2^+$~\cite{PRD57_6948,NPB,PRD92_034503,JHEP08_131,PRD88_014512,PRD87_074502,PRD93_074501,PRL118_082001,PRD97_034511,CPC46_011002},
while the calculations regarding the transitions of $J^P=1/2^+\rightarrow J^P=3/2^-$ are still very limited~\cite{PRD103_074505,PRD103_094516}.
That is because the LQCD calculations in $1/2^+\rightarrow 3/2^-$ are substantially more challenging due to the fact that the correlation functions for negative-parity baryons have more statistical noise than those for the lightest positive-parity baryons~\cite{2205.15373}. 
On the other hand, no experimental data is available to calibrate the calculations in $1/2^+\rightarrow 3/2^-$ transitions~\cite{pdg2020}.
Recently, LQCD extended the prediction on negative-parity baryons by performing the first calculation in $\Lambda^+_c\rightarrow \Lambda(1520)\ell^+\nu_{\ell}$~\cite{PRD105_L051505,PRD105_054511} decays, under the approximation that the $\Lambda(1520)$ is a stable particle under the strong interaction because of its narrow width. 
An experimental measurement of $\Lambda_c^+\rightarrow \Lambda(1520)e^+\nu_e$ will certainly provide a valuable check of the methodology applied in LQCD calculations, which is largely shared also with the calculations in 
$\Lambda_b$ decays~\cite{PRD103_074505,PRD103_094516,PRD105_054511}.
Comparison of the $\Lambda_c^+\rightarrow \Lambda(1520)/\Lambda(1405)e^+\nu_e$ decay rates between measurement and theoretical predictions provides a necessary check of the calculations from the nonrelativistic quark model~\cite{PRD95_053005} and the constituent quark model~\cite{PRC72_035201}.

In this paper, we perform the first study of the $\Lambda_c^+$ SL decay $\Lambda_c^+\rightarrow pK^- e^+\nu_e$. 
We further search for SL decays of $\Lambda_c^+\rightarrow \Lambda(1405)(\rightarrow pK^-)e^+\nu_e$ and $\Lambda_c^+\rightarrow \Lambda(1520)(\rightarrow pK^-)e^+\nu_e$, which are expected to represent the dominant components in $\Lambda_c^+\rightarrow \Lambda^* e^+\nu_e$ decays~\cite{PRC72_035201,PRD95_053005}, by investigating the $pK^-$ invariant-mass spectrum in $\Lambda_c^+\rightarrow pK^- e^+\nu_e$ data. 
The analysis is performed using data sets collected at BESIII with center-of-mass energies of $\sqrt{s}=$~4.600, 4.612, 4.628, 4.641, 4.661, 4.682, 4.699\,GeV. The total integrated luminosity for these data sets is $4.5~\mathrm{fb}^{-1}$~\cite{lum_4600,lum_new}. This is the largest data sample collected in $e^+e^-$ collisions near the $\Lambda_c^+\bar{\Lambda}_c^-$ pair production threshold.

The construction and performance of the BESIII detector are described in detail in Ref.~\cite{Ablikim:2009aa}.
A {\sc geant4}-based~\cite{geant4} Monte Carlo (MC) simulation package,
which includes the geometric description of the detector and the
detector response, is used to determine signal detection efficiencies and
to estimate potential background contributions. Signal MC samples of $e^+e^-\rightarrow \Lambda_c^+\bar{\Lambda}_c^-$ with a
$\Lambda_c^+$ baryon decaying to $pK^-e^+\nu_e$ or $\Lambda(1520)/\Lambda(1405) e^+\nu_e$  together with a
$\bar{\Lambda}_c^-$ decaying to the analyzed hadronic decay mode described below are generated by {\sc kkmc}~\cite{kkmc} 
with {\sc evtgen}~\cite{nima462_152}, with initial-state radiation (ISR)~\cite{SJNP41_466} and final-state radiation (FSR)~\cite{plb303_163} effects included.
The simulation of the SL decays $\Lambda_c^+\rightarrow pK^-e^+\nu_e$ or $\Lambda_c^+\rightarrow \Lambda(1405)/\Lambda(1520) e^+\nu_e$ is modeled with a phase-space  generator.
To study the possible peaking and combinatorial background contributions, inclusive MC samples consisting of open-charm states, radiative
return to charmonium(-like) $\psi$ states at lower masses and continuum processes of $q\bar{q}$ ($q=u,d,s$), along with Bhabha scattering, $\mu^+\mu^-$, $\tau^+\tau^-$ 
and $\gamma\gamma$ events are generated.

The first step of the analysis is the selection of ``single-tag" (ST) events with a fully
reconstructed $\bar{\Lambda}_c^-$ candidate. The $\bar{\Lambda}_c^-$ hadronic decay modes used in this analysis are:
$\bar{\Lambda}^-_c\rightarrow \bar{p} K^0_S$, $\bar{p} K^+\pi^-$,
$\bar{p}K^0_S\pi^0$, $\bar{p} K^+\pi^-\pi^0$, $\bar{p}
K^0_S\pi^+\pi^-$, $\bar{\Lambda}\pi^-$, $\bar{\Lambda}\pi^-\pi^0$,
$\bar{\Lambda}\pi^-\pi^+\pi^-$, $\bar{\Sigma}^0\pi^-$,
$\bar{\Sigma}^-\pi^0$, $\bar{\Sigma}^-\pi^+\pi^-$, $\bar{p}\pi^+\pi^-$, $\bar{\Sigma}^0\pi^+\pi^-\pi^-$, and $\bar{\Sigma}^0\pi^-\pi^0$, where the intermediate particles $K^0_S$, $\bar{\Lambda}$, $\bar{\Sigma}^0$,
$\bar{\Sigma}^-$, and $\pi^0$ are reconstructed via their decays:
$K^0_S\rightarrow \pi^+\pi^-$, $\bar{\Lambda}\rightarrow
\bar{p}\pi^+$, $\bar{\Sigma}^0\rightarrow \gamma\bar{\Lambda}$ with
$\bar{\Lambda}\rightarrow \bar{p}\pi^+$, $\bar{\Sigma}^-\rightarrow
\bar{p}\pi^0$, and $\pi^0\rightarrow \gamma\gamma$.
Within this ST sample a search is then performed for $\Lambda_c^+\rightarrow pK^-e^+\nu_e$ decay.
The events passing this selection are referred to as the double-tag (DT) sample.  For a specific tag mode $i$, the ST and DT
event yields can be written as
$$N^{i}_{\rm ST}=2N_{\bar{\Lambda}_c\Lambda_c}\mathcal{B}^i_{\rm ST}\epsilon^i_{\rm
ST}~~{\rm and}~~N^{i}_{\rm DT}=2N_{\bar{\Lambda}_c\Lambda_c}\mathcal{B}^i_{\rm
ST}\mathcal{B}_{\rm SL}\epsilon^i_{\rm DT},$$
where $N_{\bar{\Lambda}_c\Lambda_c}$ is the number of $\bar{\Lambda}_c\Lambda_c$ pairs; $\mathcal{B}^i_{\rm ST}$ and
$\mathcal{B}_{\rm SL}$ are the BFs of the $\bar{\Lambda}_c^-$ tag mode and the $\Lambda_c^+$ SL decay
mode, respectively; $\epsilon^i_{\rm ST}$ is the efficiency for finding the tag candidate; and
$\epsilon^i_{\rm DT}$ is the efficiency for simultaneously finding the tag $\bar{\Lambda}^-_c$ and the SL decay.
The BF of the SL decay can be expressed as:
\begin{equation}
\mathcal{B}_{\rm SL}=\frac{N_{\rm DT}}{\sum N^{i}_{\rm
ST}\times\epsilon^i_{\rm DT}/\epsilon^i_{\rm ST}}=\frac{N_{\rm DT}}{N_{\rm ST}\times\epsilon_{\rm SL}}, 
\label{eq:branch}
\end{equation}
where $N_{\rm DT}$ is the total yield of DT events, $N_{\rm ST}$ is the total ST yield, and
$\epsilon_{\rm SL}=\frac{\sum N^{i}_{\rm ST}\times\epsilon^i_{\rm DT}/\epsilon^i_{\rm ST}}{\sum N^{i}_{\rm ST}}$
is the average efficiency for finding a SL decay weighted by the relative yields of tag modes in data.

Selection criteria for $\gamma$, $\pi^\pm$, $K^\pm$, $p (\bar{p})$ as well as the reconstruction of $\pi^0$ and $K^0_S$ candidates are the same as those used in
Refs.~\cite{PRL115_221805,PLB767_42}.  
The invariant masses $M_{\bar{p}\pi^+}$, $M_{\gamma\bar{\Lambda}}$, and $M_{\bar{p}\pi^0}$ are required to be
within $(1.110,~1.121)$~GeV/$c^2$, $(1.179,~1.205)$~GeV/$c^2$, and $(1.171,~1.204)$~GeV/$c^2$ to form
candidates of $\bar{\Lambda}$, $\bar{\Sigma}^0$, and $\bar{\Sigma}^-$, respectively.

\begin{figure}[htbp]
\begin{center}
   \includegraphics[width=\linewidth]{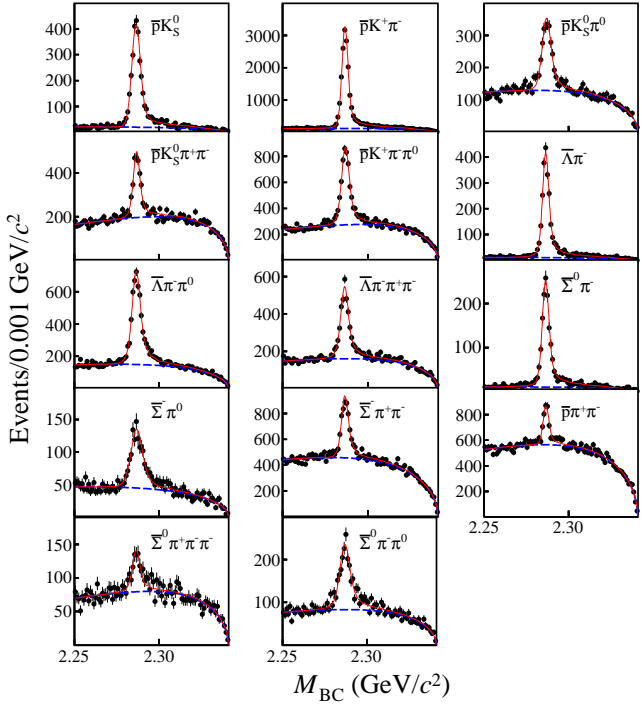}
\caption{Fits to the $M_{\rm BC}$ distributions for different ST modes at $\sqrt{s}=4.682$~GeV.
The points with error bars are data, the (red) solid curves show the total
fits and the (blue) dashed curves are the fitted backgrounds. }
\label{fig:tag_lambdac}
\end{center}
\end{figure}

The ST $\bar{\Lambda}^-_c$ signals are identified using the beam-constrained mass: 
\begin{equation}
M_{\rm BC}=\sqrt{(\sqrt{s}/2)^2/c^4-|\vec{p}_{\bar{\Lambda}^-_c}|^2/c^2},
\end{equation}
where $\vec{p}_{\bar{\Lambda}^-_c}$ is the measured momentum of the ST $\bar{\Lambda}^-_c$.
The energy difference $\Delta E=\sqrt{s}/2-E_{\bar{\Lambda}^-_c}$ is defined to improve the signal significance for ST $\bar{\Lambda}^-_c$ baryons, where $E_{\bar{\Lambda}^-_c}$ is the measured energy.
If more than one $\bar{\Lambda}_c^-$ tag is reconstructed in the event, only the tag with the minimum $|\Delta E|$ is kept to avoid double counting of STs with the same final state. The $M_{\rm BC}$ distributions at $\sqrt{s}=4.682$~GeV for the fourteen ST modes are shown in Fig.~\ref{fig:tag_lambdac}. 
An unbinned maximum-likelihood fit is performed to the spectra, using the MC-simulated signal shape convolved with a Gaussian function accounting for differences of resolutions between data and MC simulation to describe the signal and an ARGUS function~\cite{plb241_278} to describe the background. The
signal yield is determined in the  mass region $(2.28, 2.30)$~GeV/$c^2$, which is regarded as the signal region.
The $\Delta E$ requirements, the $M_{\rm BC}$ distributions for the other data sets and their ST yields are documented in the Appendix A. 
The total ST yield reconstructed in the full data sample is $N_{\rm ST}=122\,268\pm474$, where only the statistical uncertainty is reported.

Candidates from $\Lambda^+_c\rightarrow pK^-e^+\nu_e$ are
selected from the remaining particles recoiling against the ST
$\bar{\Lambda}^-_c$ candidates, with the requirement that there be exactly three charged tracks.
To select protons and kaons, the same
criteria as those used in the ST selection are used. The proton and kaon charges must be opposite in sign.
Detection and reconstruction of the positron follow the procedures in Ref.~\cite{PRL115_221805}.
Background from $\Lambda_c^+\rightarrow pK^-\pi^+$ decays is rejected by requiring the $pK^-e^+$ invariant mass ($M_{pK^-e^+}$) to be less than 2.15~GeV/$c^2$. 
To suppress contamination from $\Lambda_c^+\rightarrow pK^-\pi^+\pi^0$ decays, a search is made for an additional $\pi^0$  in the recoiling system of the $\bar{\Lambda}_c^-$ baryon.  If found, then a candidate $\Lambda_c^+\rightarrow pK^-\pi^+\pi^0$ decay is reconstructed, where the positron is now assigned a pion hypothesis.  For the event to be retained, it is required that the beam-constrained mass of this candidate falls outside the signal region.

The energy and momentum carried by the neutrino are denoted by $E_{\rm miss}$ and $\vec{p}_{\rm miss}$, respectively. They are calculated from
the energies and momenta of the tag ($E_{\bar{\Lambda}_c^-}$, $\vec{p}_{\bar{\Lambda}_c^-}$) and the measured SL decay products ($E_{\rm SL}=E_{p}+E_{K^-}+E_{e^+}$,
$\vec{p}_{\rm SL}=\vec{p}_{p}+\vec{p}_{K^-}+\vec{p}_{e^+}$) using the relations $E_{\rm miss}=\sqrt{s}/2-E_{\rm SL}$ and $\vec{p}_{\rm miss}=\vec{p}_{\Lambda_c^+}-\vec{p}_{\rm SL}$  in the initial $e^+e^-$ rest frame. 
Here, the momentum $\vec{p}_{\Lambda_c^+}$ is given by
$\vec{p}_{\Lambda_c^+}=-\hat{p}_{\rm tag}\sqrt{(\sqrt{s}/2)^2-m^2_{\bar{\Lambda}^-_c}}$, where $\hat{p}_{\rm tag}$ is the
direction of the momentum of the ST $\bar{\Lambda}^-_c$ and
$m_{\bar{\Lambda}^-_c}$ is the known $\bar{\Lambda}^-_c$
mass~\cite{pdg2020}.  
Information about the undetected neutrino is obtained by using the variable $U_{\rm miss}$:
\begin{equation}
U_{\rm miss} \equiv E_{\rm miss}-c|\vec{p}_{\rm miss}| .
\end{equation}
The $U_{\rm miss}$ distribution
is expected to peak at zero for the events of $\Lambda^+_c\rightarrow pK^-e^+\nu_e$.

\begin{figure}[tp!]
\begin{center}
   \includegraphics[width=\linewidth]{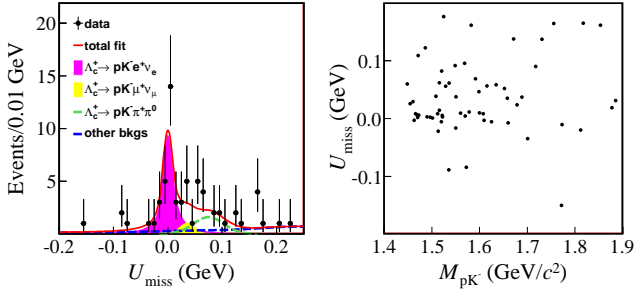}
   \caption{ (Left) Fit to the $U_{\rm miss}$ distribution for $\Lambda_c^+\rightarrow pK^- e^+\nu_e$ in data. (Right) The distribution of  $U_{\rm miss}$ {\it vs.} $M_{pK^-}$ for $\Lambda_c^+\rightarrow pK^- e^+\nu_e$ candidates. }
\label{fig:umis}
\end{center}
\end{figure}

Figure~\ref{fig:umis} (Left) shows the $U_{\rm miss}$ distribution of the reconstructed candidates for $\Lambda_c^+\rightarrow pK^-e^+\nu_e$ in data.
To obtain the signal yield, the $U_{\rm miss}$ distribution is described with four components: a signal function $f$ which consists of a Gaussian to describe the core of the 
$U_{\rm miss}$
distribution and two power-law tails to account for initial- and final-state radiation~\cite{PRD79_052010,PRL115_221805}, two MC-derived shapes describing components from 
$\Lambda_c^+\rightarrow pK^-\pi^+\pi^0$ and $\Lambda_c^+\rightarrow pK^-\mu^+\nu_{\mu}$, and an MC-derived non-resonant shape describing the combinatorial backgrounds. 
The yield of the $\Lambda_c^+\rightarrow pK^-\mu^+\nu_{\mu}$ component, $N_{\rm DT}^{pK^-\mu^+\nu_{\mu}}$,  is related to $N_{\rm DT}^{pK^-e^+\nu_e}$, which is the yield of $\Lambda_c^+\rightarrow pK^-e^+\nu_e$ decay,  by:
\begin{equation}
N_{\rm DT}^{pK^-\mu^+\nu_{\mu}}=N_{\rm DT}^{pK^-e^+\nu_e}\times R_{\epsilon}\times R_{\mathcal{B}},
\end{equation}
where $R_{\mathcal{B}}=\frac{\mathcal{B}(\Lambda^+_c\rightarrow pK^-\mu^+\nu_{\mu})}{\mathcal{B}(\Lambda^+_c\rightarrow pK^-e^+\nu_{e})}=0.88\pm0.03$~\cite{PRD95_053005}. 
The uncertainty on $R_{\mathcal{B}}$ is evaluated by comparing the difference of the BFs of $\Lambda(1405)$ and $\Lambda(1520)$ resonances decaying into $N\bar{K}e^+\nu_e$ and $N\bar{K}\mu^+\nu_{\mu}$ final states.
The parameter $R_{\epsilon}$, defined as the relative detection efficiency between $\Lambda^+_c\rightarrow pK^-\mu^+\nu_{\mu}$ and $\Lambda^+_c\rightarrow pK^-e^+\nu_{e}$, is evaluated to be $0.15$ with MC simulation. 
The event yield for $\Lambda_c^+\rightarrow pK^-e^+\nu_e$, as determined from the fit to the $U_{\rm miss}$ distribution, is $N_{\rm DT}^{pK^-e^+\nu_e}=33.5\pm6.3$, where the uncertainty is statistical.
The statistical significance of the $\Lambda_c^+\rightarrow pK^- e^+\nu_e$ signal is determined to be $8.9\sigma$, calculated via $\sqrt{-2\times \Delta {\rm ln}\mathcal{L}}$, where $\Delta {\rm ln}\mathcal{L}$ is the variation in ${\rm ln}\mathcal{L}$ of the likelihood fit with and without the signal component included.

\begin{figure}[tp!]
\begin{center}
   \includegraphics[width=\linewidth]{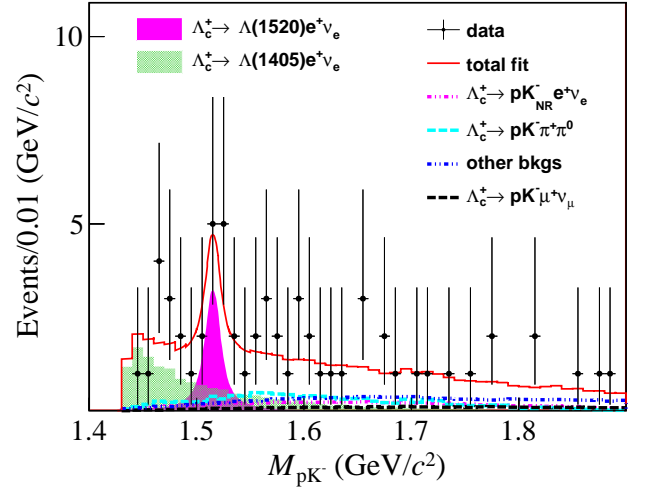}
   \caption{ Projection of the 2D-fit on the $M_{pK^-}$ axis for $\Lambda_c^+\rightarrow pK^-e^+\nu_e$ candidates in data.}
\label{fig:scat}
\end{center}
\end{figure}

To search for $\Lambda_c^+\rightarrow \Lambda(1520)/\Lambda(1405)e^+\nu_e$, the distribution of $U_{\rm miss}$ {\it vs.} $M_{pK^-}$ is studied, 
as shown in Fig.~\ref{fig:umis} (Right). 
An accumulation of events around the intersection of the $\Lambda(1520)/\Lambda(1405)$ and $pK^-e^+\nu_e$ signal regions is observed. 
To extract the yield of $\Lambda_c^+\rightarrow \Lambda(1520)/\Lambda(1405)e^+\nu_e$, a two-dimensional (2D) likelihood fit is performed to the $M_{pK^-}$ and $U_{\rm miss}$ distributions. 
For each component, the 2D distribution of $M_{pK^-}$ and $U_{\rm miss}$ is modeled with a product of two one-dimensional probability density functions (PDFs), one for each dimension. 
The signal functions in the $M_{pK^-}$ distribution for $\Lambda(1520)$ and $\Lambda(1405)$ are described by a Breit-Wigner (BW) function and a ${\rm Flatt\acute{e}}$-parameterization~\cite{PRL115_072001}, respectively. The masses and widths for $\Lambda(1520)$ and $\Lambda(1405)$ are fixed to the PDG values~\cite{pdg2020}.
The $M_{pK^-}$ distributions of the non-resonant (NR) components of $\Lambda_c^+\rightarrow (pK^-)_{\rm NR}e^+\nu_e$ and 
$\Lambda_c^+\rightarrow pK^-\mu^+\nu_{\mu}$ are described with phase-space models, while for the components from $\Lambda_c^+\rightarrow pK^-\pi^+\pi^0$ and the other background sources, the MC-derived shapes are used to describe the $M_{pK^-}$ distributions. The $U_{\rm miss}$ distributions from $\Lambda_c^+\rightarrow \Lambda(1520/1405)e^+\nu_e$ and $\Lambda_c^+\rightarrow (pK^-)_{\rm NR}e^+\nu_e$ decays are both described by the function $f$ with parameters taken from MC simulation. For the other components, the shapes obtained from MC simulation are used. 
The projection of the 2D fit on the $M_{pK^-}$ axis is shown in Fig.~\ref{fig:scat}. 
The fitted DT yields for $\Lambda_c^+\rightarrow \Lambda(1520)e^+\nu_e$ and $\Lambda_c^+\rightarrow \Lambda(1405) e^+\nu_e$ are $8.4\pm4.3$ and $14.8\pm6.7$, respectively, where the uncertainties are statistical only. 
The statistical significance of including only $\Lambda_c^+\rightarrow \Lambda(1520) e^+\nu_e$ or $\Lambda_c^+\rightarrow \Lambda(1405) e^+\nu_e$ is evaluated to be $3.8\sigma$ for both, with respect to the hypothesis that neither of the two states is included. It suggests that the two resonances contribute equally and neither of them can be neglected.

The averaged efficiencies for detecting the 
$\Lambda_c^+\rightarrow pK^-e^+\nu_e$, $\Lambda_c^+\rightarrow\Lambda(1520)(\rightarrow pK^-)e^+\nu_e$ and $\Lambda_c^+\rightarrow\Lambda(1405)(\rightarrow pK^-)e^+\nu_e$ decays are determined to be 
$\epsilon_{\rm SL}^{pK^- e^+\nu_e}=(31.01\pm 0.31)\%$, $\epsilon_{\rm SL}^{\Lambda(1520)(\rightarrow pK^-) e^+\nu_e}=(30.03\pm0.31)\%$ and $\epsilon_{\rm SL}^{\Lambda(1405)(\rightarrow pK^-)  e^+\nu_e}=(28.64\pm0.32)\%$, respectively.
Inserting the values of their DT yields, averaged efficiencies and $N^{\rm ST}$ into Eq.~(\ref{eq:branch}),  
results in $\mathcal B(\Lambda^+_c\rightarrow pK^- e^+\nu_e)=(0.88\pm0.17\pm0.07)\times10^{-3}$, $\mathcal
B(\Lambda^+_c\rightarrow \Lambda(1520)[\rightarrow pK^-]e^+\nu_e)=(0.23\pm0.12\pm0.02)\times10^{-3}$ and $\mathcal
B(\Lambda^+_c\rightarrow \Lambda(1405)[\rightarrow pK^-]e^+\nu_e)=(0.42\pm0.19\pm0.04)\times10^{-3}$, where the first uncertainties are
statistical and the second are systematic. Due to limited data, the possible interference effects between $\Lambda_c^+\rightarrow \Lambda(1520/1405)e^+\nu_e$ and 
$\Lambda_c^+\rightarrow (pK^-)_{\rm NR}e^+\nu_e$, as well as interference between $\Lambda(1520)$ and $\Lambda(1405)$ states, are ignored.

The DT technique means that the measured BFs are insensitive to any systematic uncertainties in the ST selection, as the effects of the ST selection criteria are canceled for both data and MC in the determination of the BFs.
Sources of systematic uncertainty are, instead, related to the tracking and PID efficiencies of the $e^+$ (0.4\% and 0.5\%, respectively), $p$~(1\% and 1\%), and $K^-$ (1\% and 1\%), evaluated with control samples of radiative Bhabha scattering, $J/\psi\rightarrow p\bar{p}\pi^+\pi^-$ and $J/\psi\rightarrow K^0_SK^{\mp}\pi^{\pm}$ events, respectively. 
The uncertainties arising from the fit are determined by using alternative line shapes to parameterize the signal and background contributions. 
For $\Lambda_c^+\rightarrow pK^-e^+\nu_e$, the alternative fit uses the shape from  MC simulation for the signal, with a constant to describe the combinatorial background for the $U_{\rm miss}$ distribution, which results in an uncertainty of 3.8\%.
For $\Lambda_c^+\rightarrow \Lambda(1520)/\Lambda(1405)e^+\nu_e$, the masses and widths of the signal-function are varied by $\pm 1\sigma$, and a data-driven $M_{pK^-}$ shape is used to describe the $\Lambda_c^+\rightarrow pK^-\pi^+\pi^0$ background, which results in an uncertainty of 4.6/3.9\%.
The uncertainties due to the knowledge of $R_{\mathcal{B}}$ on the three BFs are estimated to be 3.0\%, 1.4\% and 1.4\% by varying the nominal value of $R_{\mathcal{B}}$ by $\pm0.03$ in the fits.
The uncertainty of neglecting the possible decay $\Lambda_c^+\rightarrow \Lambda(1520/1405)(\rightarrow pK^-)\mu^+\nu_{\mu}$ is evaluated to be 4.8/5.4\%, by varying the description of $\Lambda_c^+\rightarrow pK^-\mu^+\nu_{\mu}$ component with $\Lambda_c^+\rightarrow (pK^-)_{\rm NR}\mu^+\nu_{\mu}$ and $\Lambda_c^+\rightarrow \Lambda(1520)/\Lambda(1405)\mu^+\nu_\mu$.
The uncertainty arising from the MC model of $\Lambda_c^+\rightarrow pK^-e^+\nu_e$ is assigned by varying the relative fraction of $\Lambda_c^+\rightarrow (pK^-)_{\rm NR}e^+\nu_e$ and $\Lambda_c^+\rightarrow \Lambda(1520)/\Lambda(1405)e^+\nu_e$ by $\pm 1\sigma$ in the MC generation. 
In the case of $\Lambda_c^+\rightarrow \Lambda(1520)/\Lambda(1405)e^+\nu_e$, a new MC model based on leading-order heavy-quark effective theory is introduced~\cite{PRD95_053005}.
These alternative models lead to a relative difference of 3.7\% and 2.7/3.6\% in the efficiencies of $\Lambda_c^+\rightarrow pK^-e^+\nu_e$ and $\Lambda_c^+\rightarrow \Lambda(1520)/\Lambda(1405)e^+\nu_e$, respectively, which is assigned as the corresponding uncertainty.
To account for neglecting possible interference effects in measuring the BF of $\Lambda_c^+\rightarrow \Lambda(1520)/\Lambda(1405)e^+\nu_e$, an additional 6.1\% uncertainty is assigned based on studies of the inclusive $K\pi$ system in $D$ SL decays~\cite{PRD94_032001,PRD99_011103}.
In addition the uncertainties of the requirements on $M_{\rm BC}$ (2.1\%), $M_{pK^-e^+}$ (3.1\%), $N_{\rm ST}$ (1.0\%) and the MC sample size (1.0\%) are considered. 
Adding these contributions in quadrature gives a total systematic uncertainty of 7.8\% for $\mathcal{B}(\Lambda_c^+\rightarrow pK^- e^+\nu_e)$, 10.4\% for $\mathcal{B}(\Lambda_c^+\rightarrow \Lambda(1520) e^+\nu_e)$ and 10.7\% for $\mathcal{B}(\Lambda_c^+\rightarrow \Lambda(1405) e^+\nu_e)$. When considering each of these systematic contributions in turn, the minimum significance for $\Lambda_c^+\rightarrow pK^- e^+\nu_e$ is determined to be $8.2\sigma$. The minimum significance for $\Lambda_c^+\rightarrow \Lambda(1520)e^+\nu_e$ and $\Lambda_c^+\rightarrow \Lambda(1405)e^+\nu_e$ are determined to be $3.3\sigma$ and $3.2\sigma$, respectively.

\normalsize
\begin{table}
\caption{Comparison of $\mathcal{B}(\Lambda_c^+\rightarrow \Lambda(1520)/\Lambda(1405)e^+\nu_e)$ [in $\times 10^{-3}$] between theoretical calculations and this measurement. The BF of 
$\Lambda(1405)\rightarrow pK^-$ is unknown~\cite{pdg2020}}.
\begin{center}
\resizebox{!}{1.15cm}{
\begin{tabular}
{l|c|c} \hline\hline 
 &  $\mathcal{B}(\Lambda_c^+\rightarrow \Lambda(1520)e^+\nu_{e})$  & $\mathcal{B}(\Lambda_c^+\rightarrow \Lambda(1405)e^+\nu_{e})$ \\ \hline
\normalsize
Constituent quark model~\cite{PRC72_035201} & 1.01 & 3.04 \\ 
Molecular state~\cite{PRD93_014021}                                & $--$ & $0.02$ \\
Nonrelativistic quark model~\cite{PRD95_053005}     & 0.60  & 2.43 \\
Lattice QCD~\cite{PRD105_L051505,PRD105_054511}     & $0.512\pm0.082$ & $--$ \\
Measurement    & $1.02\pm0.52\pm0.11$ & \normalsize $\frac{0.42\pm0.19\pm0.04}{\mathcal{B}(\Lambda(1405)\rightarrow pK^-)}$ \\
\hline\hline
\end{tabular}
}
\label{tab:theorybf}
\end{center}
\end{table}

In summary, using 4.5~fb$^{-1}$ of data collected at the centre-of-mass energies from 4.600 GeV to 4.699 GeV, the SL decay $\Lambda_c^+\rightarrow pK^-e^+\nu_e$ is observed 
with $8.2\sigma$ significance. We also find evidence for $\Lambda_c^+\rightarrow \Lambda(1520)e^+\nu_e$ and $\Lambda_c^+\rightarrow \Lambda(1405)e^+\nu_e$ with a significance of $3.3\sigma$ and $3.2\sigma$, respectively.
The measured BFs are $\mathcal B(\Lambda^+_c\rightarrow pK^- e^+\nu_e)=(0.88\pm0.17\pm0.07)\times10^{-3}$, $\mathcal
B(\Lambda^+_c\rightarrow \Lambda(1520)[\rightarrow pK^-]e^+\nu_e)=(0.23\pm0.12\pm0.02)\times10^{-3}$ and $\mathcal
B(\Lambda^+_c\rightarrow \Lambda(1405)[\rightarrow pK^-]e^+\nu_e)=(0.42\pm0.19\pm0.04)\times10^{-3}$.
Taking into account that $\mathcal{B}(\Lambda(1520)\rightarrow N\bar{K})=(45\pm1)\%$~\cite{pdg2020}, we measure $\mathcal
B(\Lambda^+_c\rightarrow \Lambda(1520)e^+\nu_e)=(1.02\pm0.52\pm0.11)\times10^{-3}$.
Comparisons of $\mathcal B(\Lambda^+_c\rightarrow \Lambda(1520)/\Lambda(1405)e^+\nu_e)$ between the measurement and predicted values from theoretical models~\cite{PRD95_053005,PRD93_014021,PRC72_035201} as well as the LQCD~\cite{,PRD105_L051505,PRD105_054511} are shown in Table~\ref{tab:theorybf}.
Our measured $\mathcal B(\Lambda^+_c\rightarrow \Lambda(1520)e^+\nu_e)$ is consistent with these theoretical calculations within two standard deviations.

Combing the BF of $\Lambda^+_c\rightarrow \Lambda(1520)e^+\nu_e$ measured in this work, $\tau_{\Lambda_c}$ and the $q^2$-integrated rate predicted by LQCD~\cite{PRD105_L051505,PRD105_054511}, we determine $|V_{cs}|=1.3\pm0.3_{\mathcal{B}}\pm0.1_{\rm LQCD}$, which is in consistent with $|V_{cs}|=0.97401(11)$ obtained assuming CKM unitarity~\cite{pdg2020} within one standard deviation. This is the first determination of $|V_{cs}|$ using data of baryonic SL decays in excited $\Lambda$ state.
Our results presented in this paper are valuable in extending the understanding of $\Lambda_c^+$ SL decays beyond the mode $\Lambda_c^+\rightarrow \Lambda \ell^+\nu_{\ell}$, and represent a significant advance in knowledge since the discovery of the $\Lambda_c^+$ more than 40 years ago. 
The observation of $\Lambda^+_c\rightarrow pK^- e^+\nu_e$ is that of the first SL 
$\Lambda_c$ decay mode that does not contain a $\Lambda$ baryon in the final state~\cite{PRD97_116015,PRC72_035201}.
In addition, the observed $pK^-$ invariant-mass spectrum in this work can provide new insights into the internal structure of excited $\Lambda$ states as well as in the study of hyperon spectroscopy~\cite{Prog120,Prog67_55,PRD93_014021,EPJC75_218,PRC92_055204}, complementary to the informations from the pentaquark searches using $\Lambda_b\rightarrow pK^-J/\psi$~\cite{PRL115_072001}. With the larger samples that BESIII expects to collect~\cite{CPC44_040001}, an amplitude analysis of the $pK^-$ mass spectrum will be performed to understand the internal structure of the contributing $\Lambda^*$ states. 

The BESIII collaboration thanks the staff of BEPCII and the IHEP computing center for their strong support. This work is supported in part by National Key R\&D Program of China under Contracts Nos. 2020YFA0406400,  2020YFA0406300; National Natural Science Foundation of China (NSFC) under Contracts Nos. 11635010, 11735014, 11835012, 11935015, 11935016, 11935018, 11961141012, 12022510, 12025502, 12035009, 12035013, 12192260, 12192261, 12192262, 12192263, 12192264, 12192265; the Chinese Academy of Sciences (CAS) Large-Scale Scientific Facility Program; Joint Large-Scale Scientific Facility Funds of the NSFC and CAS under Contract No. U1832207; CAS Key Research Program of Frontier Sciences under Contract No. QYZDJ-SSW-SLH040; 100 Talents Program of CAS; INPAC and Shanghai Key Laboratory for Particle Physics and Cosmology; ERC under Contract No. 758462; European Union's Horizon 2020 research and innovation programme under Marie Sklodowska-Curie grant agreement under Contract No. 894790; German Research Foundation DFG under Contracts Nos. 443159800, Collaborative Research Center CRC 1044, GRK 2149; Istituto Nazionale di Fisica Nucleare, Italy; Ministry of Development of Turkey under Contract No. DPT2006K-120470; National Science and Technology fund; STFC (United Kingdom); The Royal Society, UK under Contracts Nos. DH140054, DH160214; The Swedish Research Council; U. S. Department of Energy under Contract No. DE-FG02-05ER41374

\section{Appendix A The $M_{\rm BC}$ distributions for the other data sets}
\label{sec:appendixa}

\begin{table*}
\caption{ The $\Delta E$ requirements and the ST yields $N_{\rm ST}$ in each of the data sets.} \small
\begin{center}
\begin{tabular}
{|l|c|rrrrrrr|} \hline\hline $\bar{\Lambda}_c^-\rightarrow$ & $\Delta E$ (GeV)  & $N_{\rm ST}^{\rm 4600}$~~~ & $N_{\rm ST}^{\rm 4612}$~~~ & $N_{\rm ST}^{\rm 4628}$~~~ & $N_{\rm ST}^{\rm 4641}$~~~ & $N_{\rm ST}^{\rm 4661}$~~~ & $N_{\rm ST}^{\rm 4682}$~~~ & $N_{\rm ST}^{\rm 4699}$~~~ \\ \hline 
$\bar{p} K^0_S$                       & [$-$0.031, 0.033] & $1\,144\pm38$ & $230\pm17$ & $837\pm34$ & $948\pm35$ & $922\pm34$ & $2\,816\pm59$ & $791\pm31$ \\
$\bar{p} K^+\pi^-$                    & [$-$0.030, 0.039] & $6\,692\pm90$ & $1\,123\pm37$ & $5\,174\pm81$ & $5\,935\pm86$ & $5\,572\pm82$ & $16\,512\pm139$ & $4\,834\pm75$ \\
$\bar{p}K^0_S\pi^0$                & [$-$0.049, 0.052] & $622\pm42$ & $103\pm15$ & $545\pm40$ & $550\pm41$ & $568\pm40$ & $1\,649\pm62$ & $411\pm34$ \\
$\bar{p} K^0_S\pi^+\pi^-$        & [$-$0.048, 0.049] & $729\pm41$ & $145\pm18$ & $566\pm36$ & $644\pm40$ & $585\pm38$ & $1\,738\pm66$ & $555\pm38$ \\
$\bar{p} K^+\pi^-\pi^0$             & [$-$0.043, 0.051] & $1\,598\pm62$ & $275\pm24$ & $1\,163\pm54$ & $1\,319\pm62$ & $1\,295\pm55$ & $3\,943\pm97$ & $1\,077\pm50$ \\
$\bar{\Lambda}\pi^-$               & [$-$0.031, 0.034] & $878\pm30$ & $143\pm12$ & $712\pm30$ & $792\pm29$ & $730\pm29$ & $2\,254\pm50$ & $580\pm26$ \\
$\bar{\Lambda}\pi^-\pi^0$        & [$-$0.044, 0.057] & $1\,803\pm56$ & $279\pm22$ & $1\,339\pm48$ & $1\,600\pm52$ & $1\,443\pm49$ & $4\,211\pm85$ & $1\,258\pm47$ \\
$\bar{\Lambda}\pi^-\pi^+\pi^-$ & [$-$0.043, 0.045] & $1\,023\pm44$ & $199\pm18$ & $737\pm39$ & $960\pm44$ & $935\pm43$ & $2\,599\pm77$ & $710\pm39$ \\
$\bar{\Sigma}^0\pi^-$              & [$-$0.032, 0.040] & $577\pm28$ & $105\pm11$ & $424\pm24$ & $467\pm25$ & $503\pm24$ & $1\,423\pm41$ & $384\pm21$ \\
$\bar{\Sigma}^-\pi^0$              & [$-$0.050, 0.060] & $310\pm25$ & $70\pm11$ & $264\pm23$ & $282\pm26$ & $314\pm26$ & $827\pm42$ & $222\pm23$ \\
$\bar{\Sigma}^-\pi^+\pi^-$       & [$-$0.043, 0.052] & $1\,234\pm62$ & $224\pm24$ & $942\pm51$ & $1\,069\pm64$ & $938\pm53$ & $2\,941\pm96$ & $858\pm54$ \\
$\bar{p}^-\pi^+\pi^-$                 & [$-$0.040, 0.040] & $603\pm48$ & $128\pm21$ & $454\pm45$ & $490\pm48$ & $528\pm49$ & $1\,553\pm86$ & $443\pm50$ \\
$\bar{\Sigma}^0\pi^+\pi^-\pi^-$ & [$-$0.030, 0.030] & $224\pm18$ & $34\pm10$ & $150\pm22$ & $185\pm24$ & $144\pm23$ & $420\pm40$ & $133\pm23$ \\
$\bar{\Sigma}^0\pi^-\pi^0$       & [$-$0.030, 0.032] & $541\pm36$ & $102\pm15$ & $392\pm30$ & $470\pm32$ & $418\pm31$ & $1\,246\pm53$ & $437\pm30$ \\
\hline\hline
\end{tabular}
\end{center}
\label{tab:ntag}
\end{table*}

\begin{figure*}[htbp]
\begin{center}
   \begin{minipage}[t]{8.0cm}
   \includegraphics[width=\linewidth]{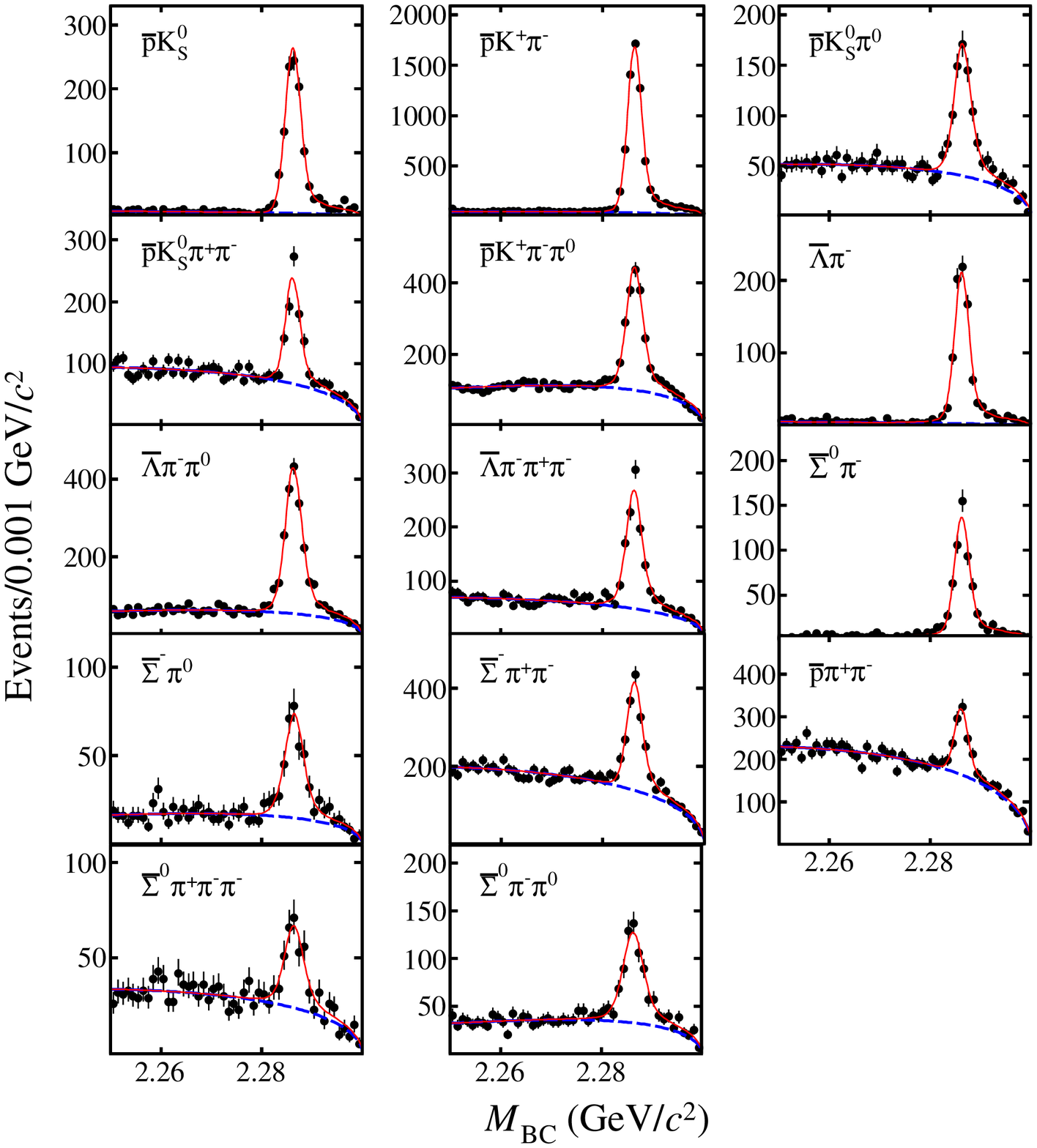}
   \end{minipage}
   \begin{minipage}[t]{8.0cm}
   \includegraphics[width=\linewidth]{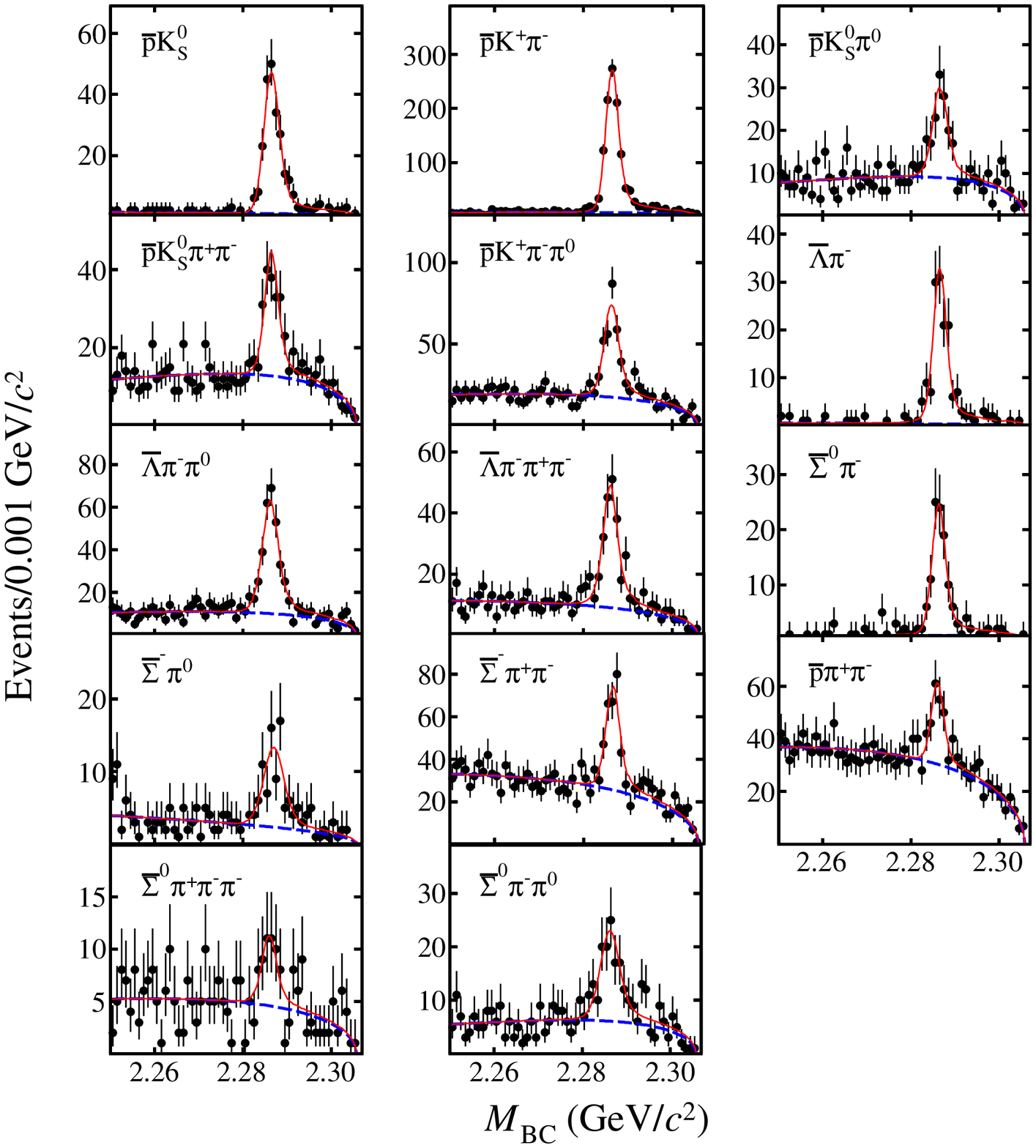}
   \end{minipage}
   \caption{Fits to $M_{\rm BC}$ distributions for different ST modes at (left) $\sqrt{s}=4.600$~GeV and (right) $\sqrt{s}=4.612$~GeV. The points with error bars are data, the (red) solid curves show the total
fits and the (blue) dashed curves are the background shapes. } 
   \label{fig:MBC4600}
\end{center}
\end{figure*}

\begin{figure*}[htbp]
\begin{center}
   \begin{minipage}[t]{8.0cm}
   \includegraphics[width=\linewidth]{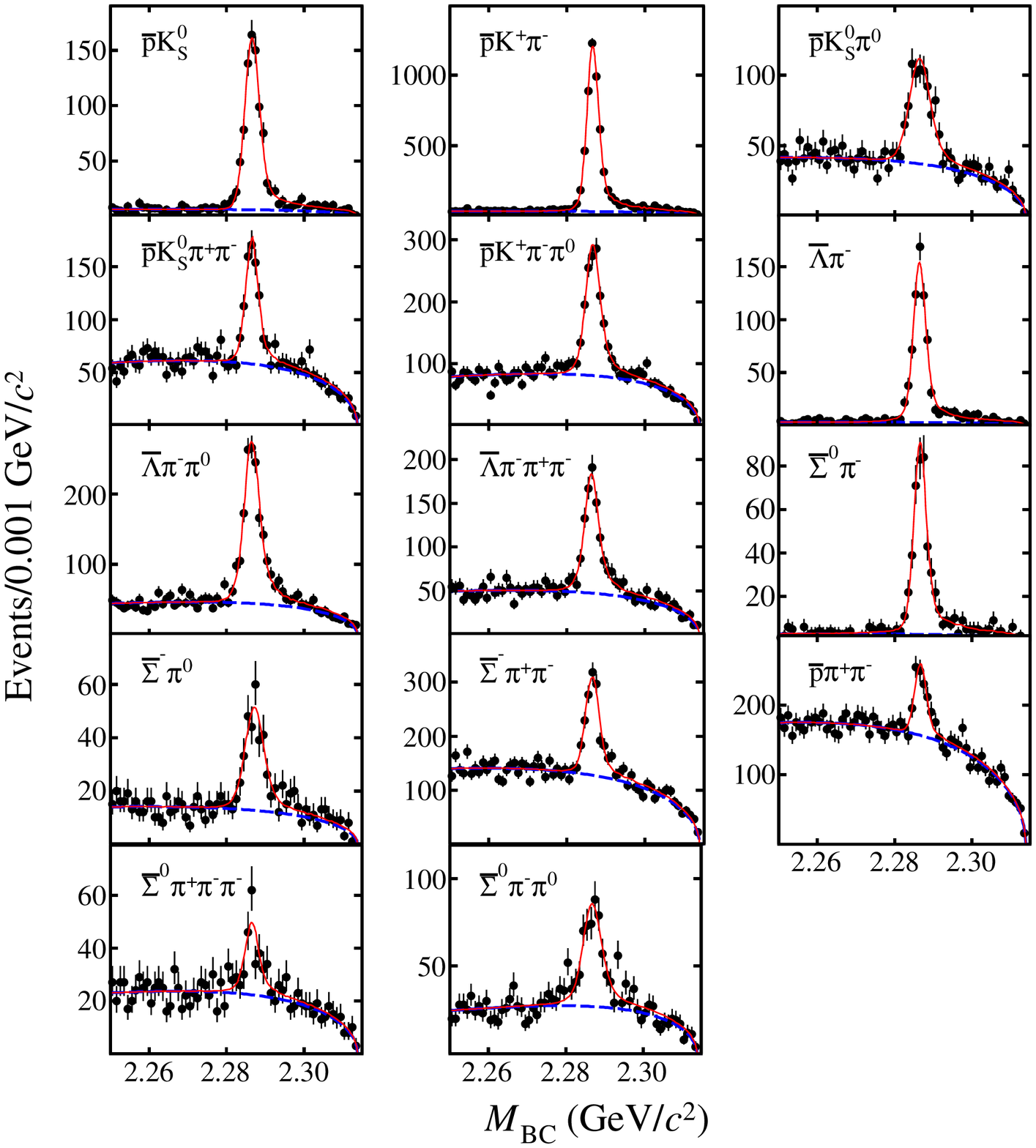}
   \end{minipage}
   \begin{minipage}[t]{8.0cm}
   \includegraphics[width=\linewidth]{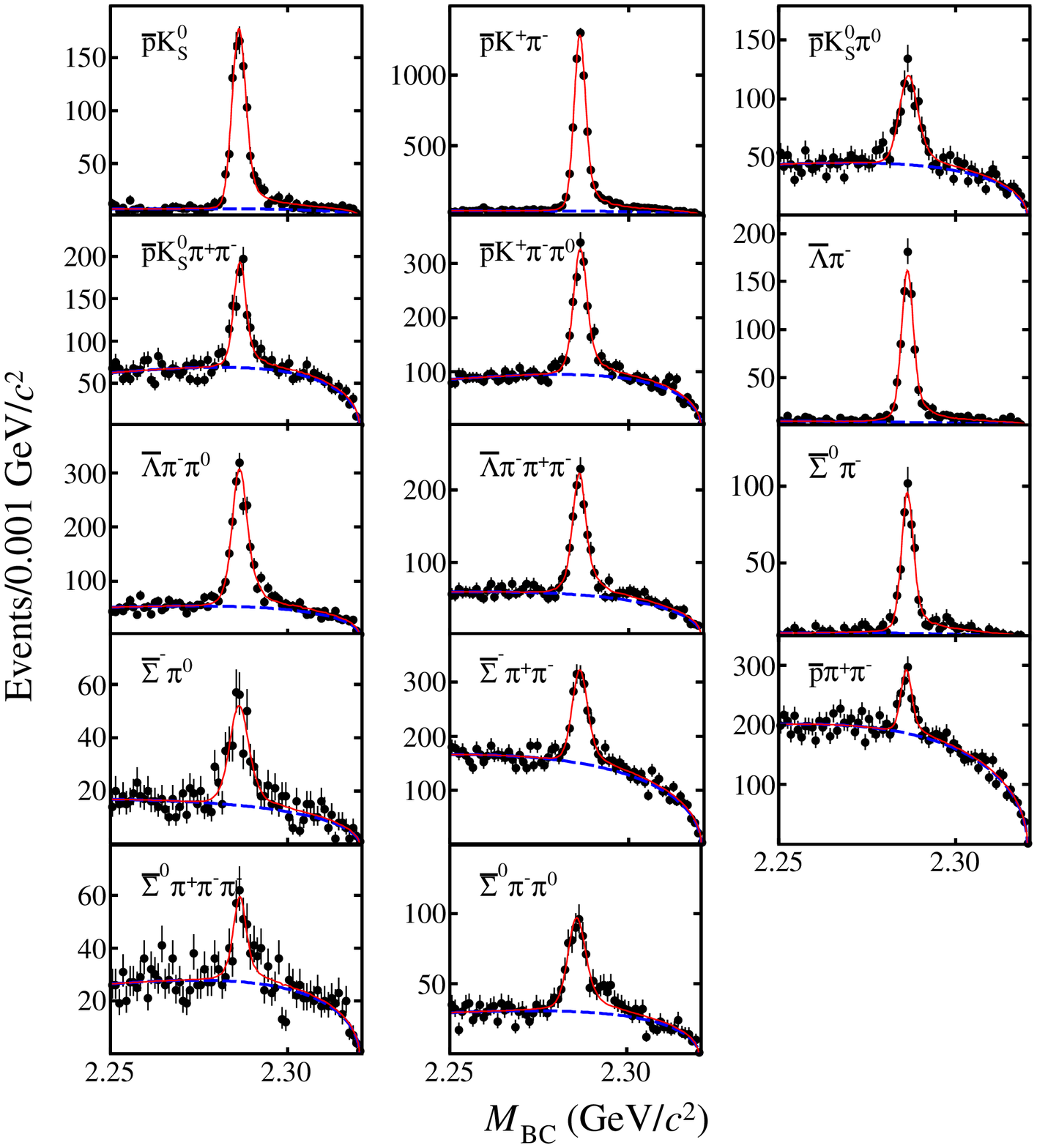}
   \end{minipage}
   \caption{Fits to $M_{\rm BC}$ distributions for different ST modes at (left) $\sqrt{s}=4.628$~GeV and (right) $\sqrt{s}=4.641$~GeV. The points with error bars are data, the (red) solid curves show the total
fits and the (blue) dashed curves are the background shapes. } 
   \label{fig:MBC4626}
\end{center}
\end{figure*}

\begin{figure*}[htbp]
\begin{center}
   \begin{minipage}[t]{8.0cm}
   \includegraphics[width=\linewidth]{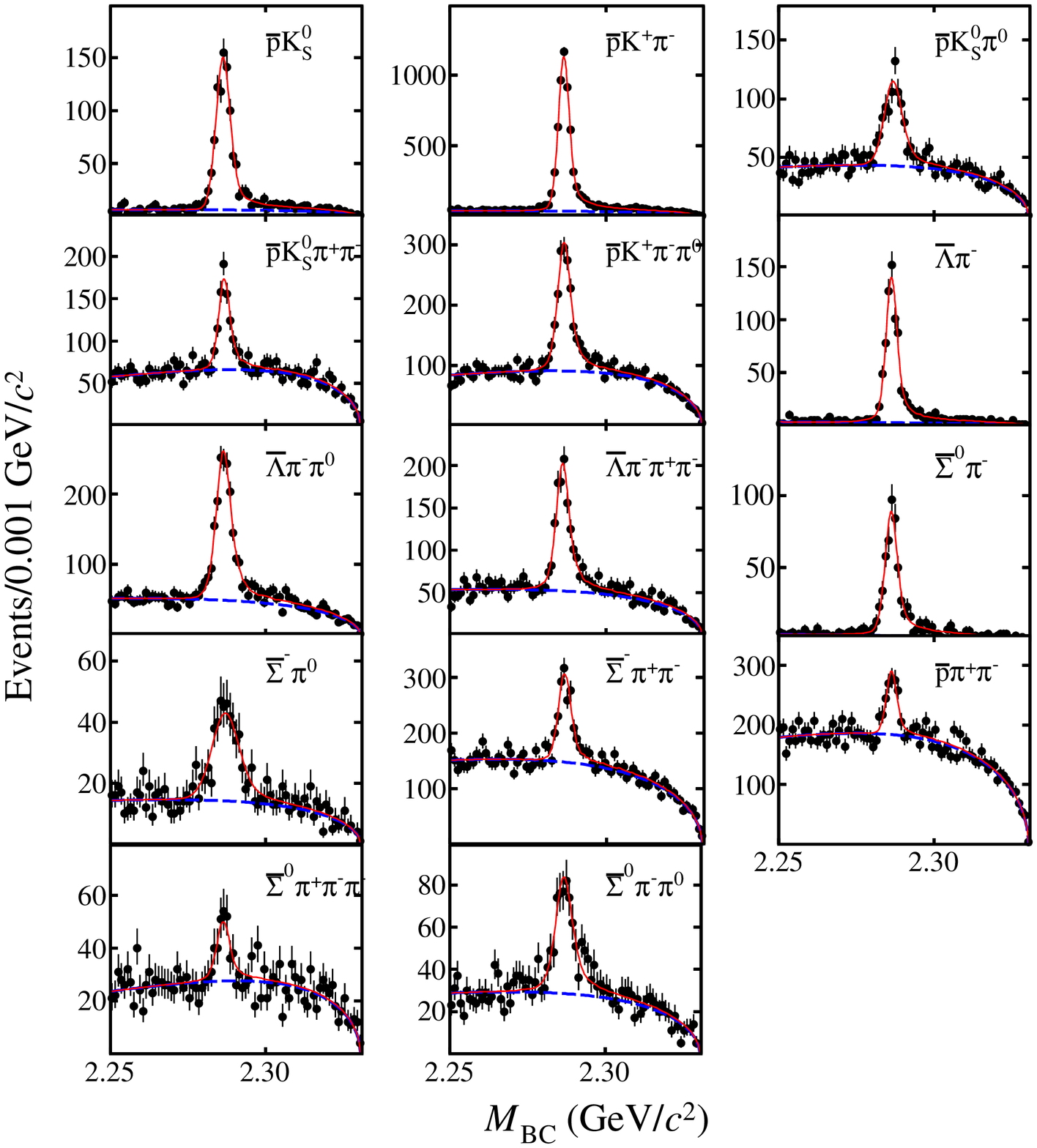}
   \end{minipage}
   \begin{minipage}[t]{8.0cm}
   \includegraphics[width=\linewidth]{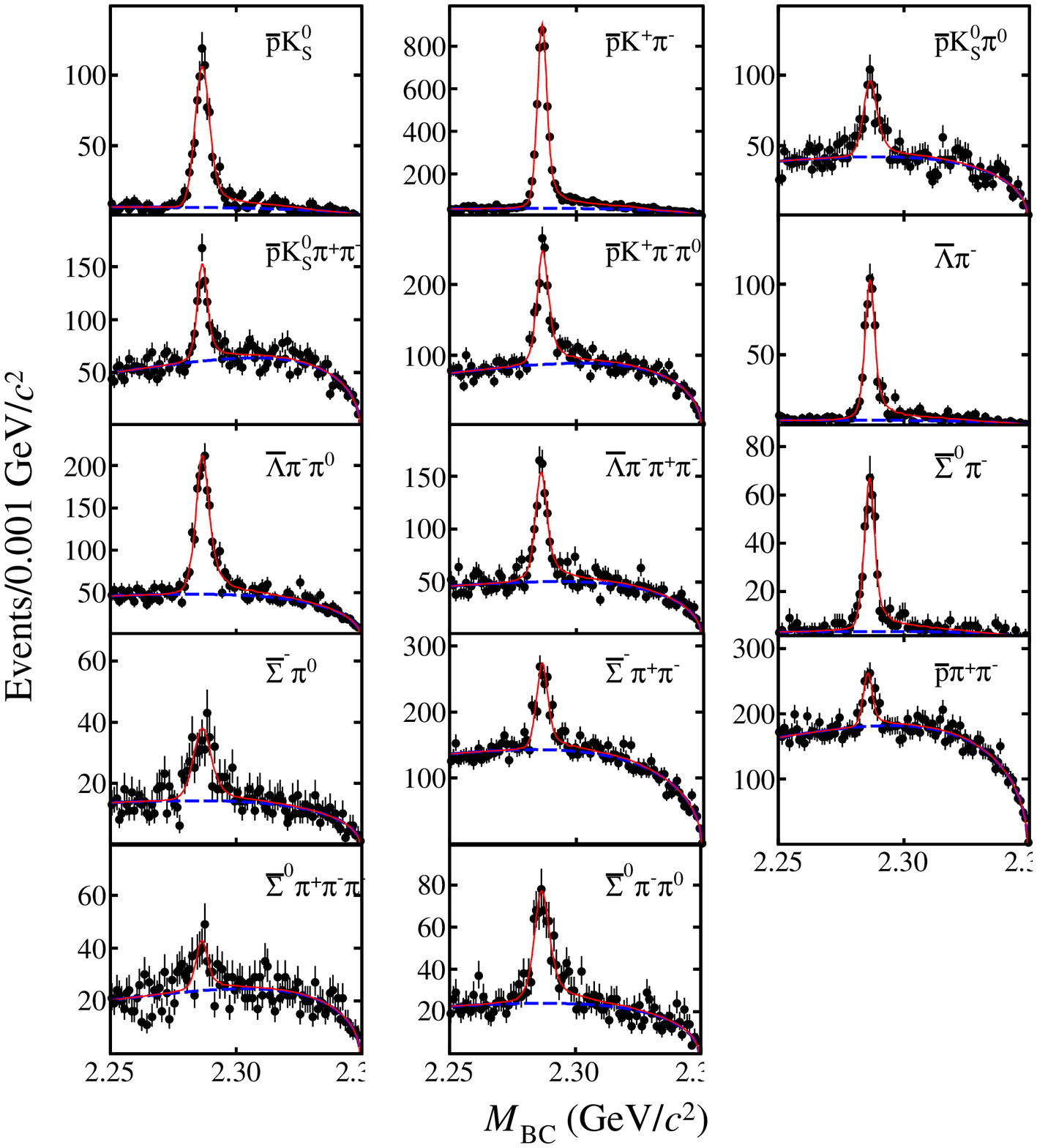}
   \end{minipage}
   \caption{Fits to $M_{\rm BC}$ distributions for different ST modes at (left) $\sqrt{s}=4.661$~GeV and (right) $\sqrt{s}=4.699$~GeV. The points with error bars are data, the (red) solid curves show the total
fits and the (blue) dashed curves are the background shapes. } 
   \label{fig:MBC4660}
\end{center}
\end{figure*}

Figures~\ref{fig:MBC4600}, \ref{fig:MBC4626} and \ref{fig:MBC4660} show the $M_{\rm BC}$ distributions obtained at $\sqrt{s}=4.600$, 4.612, 4.628, 4.641, 4.661, and 4.699 GeV, respectively. The ST yields for each of the ST modes collected at different energy points are given in Table~\ref{tab:ntag}. 

\begin{thebibliography}{99}
\bibitem{CKM} M. Kobayashi and T. Maskawa, \href{https://doi.org/10.1016/j.nuclphysbps.2005.01.049}{Prog. Theor. Phys. {\bf 49}, 652 (1973)}.
\bibitem{pdg2020} P.A. Zyla $et al.$ [Particle Data Group], \href{https://pdg.lbl.gov/}{Prog. Theor. Exp. Phys. 2020, 083C01 (2020) and 2021 update}. 

\bibitem{PRL115_221805} M.~Ablikim $et~al.$ [BESIII Collaboration], \href{https://doi.org/10.1103/PhysRevLett.115.221805}{Phys. Rev. Lett. {\bf 115}, 221805 (2015)}.
\bibitem{PLB767_42} M.~Ablikim $et~al.$ [BESIII Collaboration], \href{https://doi.org/10.1016/j.physletb.2017.01.047}{Phys. Lett. B {\bf 767}, 42 (2017)}.
\bibitem{PRL121_251801} M.~Ablikim $et~al.$ [BESIII Collaboration],  \href{https://doi.org/10.1103/PhysRevLett.121.251801}{Phys. Rev. Lett. {\bf 121}, 251801 (2018)}. 
\bibitem{2103.07064} Q.~A.~Zhang, H.~Hua, F.~Huang, R.~Li, Y.~Li, C.~D.~L{\"u}, P.~Sun, W.~Wang and Y.~B.~Yang, \href{https://doi.org/10.1088/1674-1137/ac2b12}{Chin. Phys. C {\bf 46}, 011002 (2022)}.
\bibitem{PRD104_013005} Y.~S.~Li, X.~Liu and F.~S.~Yu, \href{https://doi.org/10.1103/PhysRevD.104.013005}{Phys. Rev. D {\bf 104}, 013005 (2021)}.

\bibitem{PRC72_035201} M. Pervin, W. Roberts and S. Capstick, \href{https://doi.org/10.1103/PhysRevC.72.035201}{Phys. Rev. C {\bf 72}, 035201 (2005)}. 
\bibitem{PRD93_014021} N. Ikeno, and E. Oset, Phys. \href{https://doi.org/10.1103/PhysRevD.93.014021}{Rev. D {\bf 93}, 014021 (2016)}. 
\bibitem{PRD95_053005} M.~M.~Hussain and W.~Roberts, \href{https://doi.org/10.1103/PhysRevD.95.053005}{Phys. Rev. D {\bf 95}, 053005 (2017)}; \href{https://doi.org/10.1103/PhysRevD.95.099901}{Phys. Rev. D {\bf 95}, 099901(E) (2017)}.
\bibitem{PRD97_116015} M. Gronau and J. L. Rosner, \href{https://doi.org/10.1103/PhysRevD.97.116015}{Phys. Rev. D {\bf 97}, 116015 (2018)}.
\bibitem{PRD105_L051505} S.~Meinel and G.~Rendon, \href{https://doi.org/10.1103/PhysRevD.105.L051505}{Phys. Rev. D {\bf 105}, L051505 (2022)}.
\bibitem{PRD105_054511} S.~Meinel and G.~Rendon, \href{https://doi.org/10.1103/PhysRevD.105.054511}{Phys. Rev. D {\bf 105}, 054511 (2022)}.
\bibitem{PRC92_055204} K.~Miyahara, T.~Hyodo, E.~Oset, \href{https://doi.org/10.1103/PhysRevC.92.055204}{Phys. Rev. C {\bf 92}, 055204 (2015)}.



\bibitem{PRL2_425} R.~H.~Dalitz, S.~F.~Tuan, \href{https://doi.org/10.1103/PhysRevLett.2.425}{Phys. Rev. Lett. {\bf 2} 425 (1959)}.
\bibitem{AnnPhys}  R.~H.~Dalitz and S.~F.~Tuan, \href{https://doi.org/10.1016/0003-4916(60)90001-4}{Ann. Phys. (N.Y.) {\bf 10}, 307 (1960)}.
\bibitem{Prog120} T.~Hyodo, M.~Niiyama, \href{https://doi.org/10.1016/j.ppnp.2021.103868}{Prog. Part. Nucl. Phys. {\bf 120}103868 (2021)}.
\bibitem{EPJST} M.~Mai, \href{https://doi.org/10.1140/epjs/s11734-021-00144-7}{Eur. Phys. J. Spec. Top. {\bf 230} 1593 (2021)}.


\bibitem{PR153_1617} R.~H.~Dalitz, T.~C.~Wong, and G.~Rajasekaran, \href{https://doi.org/10.1103/PhysRev.153.1617}{Phys. Rev. {\bf 153}, 1617 (1967)}.

\bibitem{PRD57_6948} K.~C.~Bowler, R.~D.~Kenway, L.~Lellouch, J.~Nieves, O.~Oliveira, D.~G.~Richards, C.~T.~Sachrajda, N.~Stella and P.~Ueberholz, \href{https://doi.org/10.1103/PhysRevD.57.6948}{Phys. Rev. D {\bf 57}, 6948 (1998)}.
\bibitem{NPB} S.~Gottlieb and S.~Tamhankar, \href{https://doi.org/10.1016/S0920-5632(03)01612-8}{Nucl. Phys. B, Proc. Suppl. {\bf 119} 644 (2003)}.
\bibitem{JHEP08_131} A.~Datta, S.~Kamali, S. Meinel and A.~Rashed, \href{https://doi.org/10.1007/JHEP08(2017)131}{JHEP {\bf 08} (2017) 131}.

\bibitem{PRD92_034503} W.~Detmold, C.~Lehner and S.~Meinel, \href{https://doi.org/10.1103/PhysRevD.92.034503}{Phys. Rev. D {\bf 92}, 034503 (2015)}.
\bibitem{PRD88_014512} W.~Detmold, C.-J.~David~Lin, S.~Meinel and M.~Wingate, \href{https://doi.org/10.1103/PhysRevD.88.014512}{Phys. Rev. D {\bf 88}, 014512 (2013)}.
\bibitem{PRD87_074502} W.~Detmold, C.-J.~David~Lin, S.~Meinel and M.~Wingate, \href{https://doi.org/10.1103/PhysRevD.87.074502}{Phys. Rev. D {\bf 87}, 074502 (2013)}.
\bibitem{PRD93_074501} W.~Detmold, and S.~Meinel, \href{https://doi.org/10.1103/PhysRevD.93.074501}{Phys. Rev. D {\bf 93}, 074501 (2016)}.
\bibitem{PRL118_082001} S.~Meinel, \href{https://doi.org/10.1103/PhysRevLett.118.082001}{Phys. Rev. Lett. {\bf 118}, 082001 (2017)}. 
\bibitem{PRD97_034511} S.~Meinel, \href{https://doi.org/10.1103/PhysRevD.97.034511}{Phys. Rev. D {\bf 97}, 034511 (2018)}.
\bibitem{CPC46_011002} Q.~A.~Zhang, H.~Hua, F.~Huang, R.~Li, Y.~Li, C.~D.~L{\"u}, P.~Sun, W.~Wang and Y.~B.~Yang, \href{https://doi.org/10.1088/1674-1137/ac2b12}{Chin. Phys. C {\bf 46}, 011002 (2022)}.

\bibitem{PRD103_074505} S.~Meinel and G.~Rendon, \href{https://doi.org/10.1103/PhysRevD.103.074505}{Phys. Rev. D {\bf 103}, 074505 (2021)}.
\bibitem{PRD103_094516} S.~Meinel and G.~Rendon, \href{https://doi.org/10.1103/PhysRevD.103.094516}{Phys. Rev. D {\bf 103}, 094516 (2021)}.
\bibitem{2205.15373} P.~A.~Boyle $et~al.$, \href{https://doi.org/10.48550/arXiv.2205.15373}{arXiv:2205.15373}.


\bibitem{lum_4600} M.~Ablikim $et~al.$ [BESIII Collaboration], \href{https://doi.org/10.1088/1674-1137/39/9/093001}{Chin. Phys. C {\bf 39}, 093001 (2015)}.
\bibitem{lum_new} M.~Ablikim $et~al.$ [BESIII Collaboration], \href{https://doi.org/10.48550/arXiv.2205.04809}{arXiv:2205.04809}.


\bibitem{Ablikim:2009aa} M.~Ablikim $et~al.$ [BESIII Collaboration], \href{https://doi.org/10.1016/j.nima.2009.12.050}{Nucl.\ Instrum.\ Meth.\ A {\bf 614}, 345 (2010)}.
\bibitem{geant4} S.~Agostinelli $et~al.$ [GEANT4 Collaboration], \href{http://dx.doi.org/10.1016/S0168-9002(03)01368-8}{Nucl. Instrum. Meth. A {\bf 506}, 250 (2003)}.
\bibitem{kkmc} S. Jadach, B. F. L. Ward and Z. Was, \href{https://doi.org/10.1016/S0010-4655(00)00048-5}{Comput. Phys. Commun. {\bf 130}, 260 (2000)}; \href{https://doi.org/10.1103/PhysRevD.63.113009}{Phys. Rev. D {\bf 63}, 113009 (2001)}.
\bibitem{nima462_152} D.~J.~Lange, \href{https://doi.org/10.1016/S0168-9002(01)00089-4}{Nucl. Instrum. Meth. A {\bf 462}, 152 (2001)}; \href{https://doi.org/10.1088/1674-1137/32/8/001}{R. G. Ping, Chin. Phys. C {\bf 32}, 599 (2008)}.
\bibitem{SJNP41_466} E.~A.~Kurav and V.~S.~Fadin, Sov. J. Nucl. Phys. {\bf 41}, 466 (1985).
\bibitem{plb303_163} E.~Richter-Was, \href{https://doi.org/10.1016/0370-2693(93)90062-M}{Phys. Lett. B {\bf 303}, 163 (1993)}.

\bibitem{plb241_278} H. Albrecht $et~al.$ [ARGUS Collaboration], \href{https://doi.org/10.1016/0370-2693(90)91293-K}{Phys. Lett. B. {\bf 241}, 278 (1990)}.

\bibitem{PRD79_052010} J.~Y.~Ge $et~al.$ [CLEO Collaboration], \href{https://doi.org/10.1103/PhysRevD.79.052010}{Phys. Rev. D {\bf 79}, 052010 (2009)}.
\bibitem{PRL115_072001} R.~Aaij $et~al.$ [LHCb Collaboration], \href{https://doi.org/10.1103/PhysRevLett.115.072001}{Phys. Rev. Lett. {\bf 115}, 072001 (2015)}.

\bibitem{PRD94_032001} M.~Ablikim $et~al.$ [BESIII Collaboration], \href{https://doi.org/10.1103/PhysRevD.94.032001}{Phys. Rev. D {\bf 94}, 032001 (2016)}.
\bibitem{PRD99_011103} M.~Ablikim $et~al.$ [BESIII Collaboration], \href{https://doi.org/10.1103/PhysRevD.99.011103}{Phys. Rev. D {\bf 99}, 011103R (2019)}.

\bibitem{Prog67_55} T.~Hyodo, D.~Jido, \href{https://doi.org/10.1016/j.ppnp.2011.07.002}{Prog. Part. Nucl. Phys. {\bf 67}, 55 (2012)}.
\bibitem{EPJC75_218} L.~Roca, M.~Mai, E.~Oset. Ulf-G.~Meißner, \href{https://doi.org/10.1140/epjc/s10052-015-3438-1}{Eur. Phys. J. C {\bf 75}, 218 (2015)}.
\bibitem{CPC44_040001} M.~Ablikim $et~al.$ [BESIII Collaboration], \href{https://doi.org/10.1088/1674-1137/44/4/040001}{Chin. Phys. C {\bf 44}, 040001 (2020)}.

\end{thebibliography}
\end{document}